\documentclass{mn2e}

\usepackage{amsfonts}
\usepackage{amsmath}
\usepackage{graphicx}

\newcommand{\p}{\ensuremath{\partial}}
\newcommand{\del}{\ensuremath{\delta}}
\newcommand{\Del}{\ensuremath{\Delta}}
\newcommand{\lam}{\ensuremath{\lambda}}
\newcommand{\sig}{\ensuremath{\sigma}}
\newcommand{\ep}{\ensuremath{\epsilon}}
\newcommand{\lamint}[1]{\ensuremath{\int_{-\infty}^{\infty}{\frac{d\lam_{#1}}{2\pi}}}} 
\newcommand{\avg}[1]{\ensuremath{\langle \,#1\, \rangle}}
\newcommand{\etal}{\emph{et al}.}
\newcommand{\delc}{\ensuremath{\delta_c}}

\newcommand{\eqn}[1]{equation~\eqref{#1}}

\newcommand{\fig}[1]{Figure~\ref{#1}}
\newcommand{\figs}[1]{Figures~\ref{#1}}
\newcommand{\ph}[1]{\phantom{#1}}

\newcommand{\be}{\begin{equation}}
\newcommand{\ee}{\end{equation}}

\title[Excursion set with correlated steps]
      {Halo abundances and counts-in-cells:  
       The excursion set approach with correlated steps}

\author[A. Paranjape, T. Y. Lam \& R. K. Sheth]
{Aseem Paranjape$^{1}$\thanks{E-mail: aparanja@ictp.it, tszyan.lam@ipmu.jp}, 
 Tsz Yan Lam$^{2}$ \& Ravi K. Sheth$^{1,3}$\\
 $^1$ The Abdus Salam International Center for Theoretical Physics, 
      Strada Costiera 11, 34151 Trieste, Italy\\
 $^2$ IPMU, University of Tokyo, Kashiwa, Chiba 277-8583, Japan\\
 $^3$ Center for Particle Cosmology, University of Pennsylvania, 
      209 S. 33rd St., Philadelphia, PA 19104, USA}
\begin{document}
\pagerange{\pageref{firstpage}--\pageref{lastpage}}

\maketitle 

\label{firstpage}

\begin{abstract}
The Excursion Set approach has been used to make predictions for 
a number of interesting quantities in studies of nonlinear 
hierarchical clustering.  These include the halo mass function, 
halo merger rates, halo formation times and masses, halo clustering, 
analogous quantities for voids, and the distribution of dark matter 
counts in randomly placed cells.  The approach assumes that all these 
quantities can be mapped to problems involving the first crossing 
distribution of a suitably chosen barrier by random walks.  
Most analytic expressions for these distributions ignore the fact 
that, although different $k$-modes in the initial Gaussian field are 
uncorrelated, this is not true in real space:  the values of the density 
field at a given spatial position, when smoothed on different 
real-space scales, are correlated in a nontrivial way.  As a result, 
the problem is to estimate first crossing distribution by random walks 
having correlated rather than uncorrelated steps. In 1990, Peacock \&
Heavens presented a simple approximation for the first crossing
distribution of a single barrier of constant height by walks with
correlated steps.  We show that their approximation can be thought of
as a correction to the distribution associated with what we call smooth 
completely correlated walks. We then use this insight to extend their 
approach to treat moving barriers, as well as walks that are constrained 
to pass through a certain point before crossing the barrier.  
For the latter, we show that a simple rescaling, inspired by bivariate
Gaussian statistics, of the unconditional first crossing distribution,
accurately describes the conditional distribution, independently of the
choice of analytical prescription for the former.
In all cases, comparison with Monte-Carlo solutions of the problem 
shows reasonably good agreement.  This represents the first explicit 
demonstration of the accuracy of an analytic treatment of all these 
aspects of the correlated steps problem. While our main focus is on
first crossing distributions of deterministic barriers by random
walks, in an Appendix we also discuss several issues that arise upon
introducing a stochasticity in the barrier height, a topic which has
gained interest recently with regards the mapping between first
crossing distributions and halo mass functions.  
\end{abstract}


\section{Introduction}
The abundance of clusters and its evolution is a useful probe of 
the primordial fluctuation field, the subsequent expansion history 
of the universe, and the nature of gravity.  This is, in part, because 
there is an analytic framework for understanding how cluster formation 
and evolution depends on the background cosmological model.  
Analyses based on the assumption that clusters form from a spherical 
collapse suggest that a cluster today is a region that is about 200 
times the background density, and it formed from the collapse of a 
sufficiently overdense region in the initial conditions
(Gunn \& Gott 1972).  Numerical simulations suggest that this 
expectation is reasonably accurate.  

Hence, the problem of estimating cluster abundances at any given time 
reduces to the problem of estimating the abundance of sufficiently 
overdense regions in the initial conditions (Press \& Schechter 1974).  
However, the overdensity associated with a given position in space 
depends on scale (in homogeneous cosmologies, the likely range of 
overdensities is smaller on large scales).  So, to estimate cluster 
abundances, the problem is to find those regions in the initial 
conditions which are sufficiently overdense on a given smoothing scale, 
but not on a larger scale.  This is because, if the larger region is 
sufficiently overdense, then, as it pulls itself together against the 
expansion of the background universe and collapses, it will also squeeze 
the regions within it to smaller and smaller sizes.  The framework 
for not double-counting the smaller overdense regions that are embedded 
in larger ones is known as the Excursion Set approach (Epstein 1983; 
Bond \etal\ 1991; Lacey \& Cole 1993).  With some care, one can build 
an Excursion Set model for voids as well (Sheth \& van de Weygaert
2004). 

Within the context of the Excursion Set approach, the spherical evolution 
model exhibits an important technical simplification:  the critical 
overdensity \delc\ required for collapse at a given time is 
{\em independent} of the mass or size of the final object.  
In fact, neither clusters nor voids are spherical, and the critical 
overdensity associated with collapsed objects depends on how different 
from spherical the object is.  As a result, the critical overdensity 
required for collapse, when averaged over objects of a given mass, 
becomes mass dependent (Sheth, Mo \& Tormen 2001).  In modified gravity 
models, mass dependence of the critical overdensity appears even in 
spherical evolution models (Martino, Stabenau \& Sheth 2009; 
Brax, Rosenfeld \& Steer 2010).  

While mass-dependence of \delc\ does not complicate the logical 
framework of the Excursion Set description, it does impact the 
ability to obtain exact analytic expressions for the quantities of 
interest.  Nevertheless, simple accurate approximations have been 
developed (Sheth \& Tormen 2002; Lam \& Sheth 2009).  
In addition to allowing one to predict cluster abundances, these 
allow one to produce accurate Excursion Set models which describe 
how the probability distribution function of mass in randomly placed 
cells depends on cell size (Sheth 1998; Lam \& Sheth 2008).  In some 
respects, clusters can be thought of as cells of vanishingly small size, 
so the excursion set model for the counts-in-cells distribution is also 
a model for the density run around clusters (and voids) on scales that 
are larger than the virial radius (or void wall).  

To obtain analytic expressions, all of these analyses assume that 
the density field on one scale is trivially correlated with that 
on another scale.  If one plots the overdensity as a function of 
smoothing scale, then this resembles a random walk -- the usual 
assumption is that successive steps in the walk are independent of 
the previous ones.  This is known to be a bad approximation, but 
because there are no known exact solutions to the case of realistic 
correlations, the assumption of uncorrelated steps has been routine.  
This is despite the fact that Peacock \& Heavens (1990) showed how to 
derive a reasonably accurate expression for the spherical collapse 
problem and correlated steps.  However, it has received little use, 
presumably because it is only an approximation, whereas the 
corresponding problem of spherical collapse with uncorrelated steps 
was solved exactly shortly after their paper appeared (Bond \etal\ 1991).  
Bond \etal. also described a simple numerical solution to the correlated 
steps problem; it too has received little attention.  

Recently, however, there has been renewed interest in the correlated 
steps problem:  Maggiore \& Riotto (2010a) have introduced field theory 
techniques to address this problem.  In essence, this approach aims to 
solve analytically the same path integrals that Bond \etal. solved 
numerically.  This technical machinery is too complicated to solve 
exactly, but, at the end of the day, it does provide a simple analytic 
approximation for cluster abundances in the spherical collapse model.  
There is as yet no similarly simple expression for the case in which 
\delc\ is mass-dependent (although see De Simone, Maggiore \& Riotto
2011a for a treatment of correlations induced by non-Gaussian initial
conditions).

The present paper is motivated by the fact that the field theoretic 
approach yields approximate rather than exact expressions.  So it is 
interesting to ask how it compares to the older Peacock-Heavens 
approximation.  In Section~\ref{constant}, we show that the older 
approximation is, in fact, the more accurate of the two, for the case
of a constant, deterministic barrier (see below for a discussion of
the case when the barrier height is stochastic).
Therefore, the main goal of the current paper is to show how the
analysis of Peacock \& Heavens can be extended to the case of
mass-dependent $\delta_c$.  Section~\ref{moving} shows that this can
be done almost trivially.  Section~\ref{constrained} discusses an
ansatz which relates the shape of the first crossing distribution of
walks that are constrained to pass through a certain point before
first crossing the barrier to the shape of the unconditional first
crossing distribution. We combine this with the Peacock-Heavens
approach to provide a rather accurate approximation of the conditional
distribution.   
A final section summarizes and discusses some implications.  
Appendix~\ref{phMatrix} provides an alternative derivation of the 
Peacock-Heavens ansatz which yields some insight into the nature of 
their approximation, while Appendix~\ref{Wth} contains technical
details about smoothing windows. Appendix~\ref{stoc} includes a
discussion of  problems in which the barrier height is stochastic, an
issue which has gained considerable recent interest (Maggiore \& 
Riotto 2010b, Corasaniti \& Achitouv 2011b). We argue that these
latter treatments correspond to making some specific technical choices
which are difficult to test, and that there are in fact several other
(testable) options when dealing with stochastic barriers, which remain
to be explored.

In a related paper, we address the question of halo bias 
(Paranjape \& Sheth 2011), which is associated with a certain limit 
of our solution of the constrained walks problem.  
And in a third, we address the question of voids 
(Paranjape, Lam \& Sheth 2011):  here, the problem is the generalization 
of the Peacock-Heavens approximation to the case of two absorbing barriers 
rather than just one.  This also works rather well.  All of these analyses 
assume the initial fluctuation field was Gaussian.  The case of 
non-Gaussian initial conditions is discussed by Musso \& Paranjape
(2011). 

\section{The constant barrier problem}\label{constant}
For what follows, it will be useful to define 
\begin{equation}
 \sigma_j^2(R) \equiv 
 \int \frac{{\rm d}k}{k}\,\frac{k^3P(k)}{2\pi^2}\,k^{2j}\,W^2(kR)\,,
\label{sig2jdef}
\end{equation}
where $P(k)$ is the power spectrum of initial density fluctuations, 
(linearly extrapolated to present epoch) and $W$ is a smoothing
filter.  The quantity $\sigma_0^2(R)$ measures  the variance in the
field on scale $R$. We will reserve the symbol $s$ (or $S$) to denote
the variance, 
\be
s \equiv \sig_0^2(R)\,.
\label{Sdef}
\ee
We will also make use of the combination 
\begin{equation}
 \gamma \equiv \frac{\sigma_1^2}{\sigma_0\sigma_2}.
\end{equation}
For $P(k)\propto k^n$ and a Gaussian filter $W(kR)=e^{-(kR)^2/2}$,
which we will use to illustrate many of our results,
$\sigma_j^2\propto R^{-3-n-2j}$ and $\gamma^2 = (3+n)/(5+n)$.   

The Excursion set ansatz relates the abundance of halos of mass $m$ 
to the fraction of random walks which first cross a barrier of height 
$\delta_c$ on scale $s(R)$, where $m = \bar\rho\,4\pi R^3/3$.  If
$f(s)\,{\rm d}s$ denotes this fraction, then 
\begin{equation}
 \frac{m}{\bar\rho}\frac{{\rm d}n(m)}{{\rm d}m} {\rm d}m = f(s)\,{\rm
   d}s. 
 \label{ansatz}
\end{equation} 
We argue elsewhere that this relation between halo abundances and 
the first crossing distribution is not the full story.  
In what follows, we are mainly interested in making accurate estimates 
of the first crossing distribution.

\subsection{Numerical (Monte-Carlo) solution:  TopHat smoothing and
  $\Lambda$CDM $P(k)$}
\label{CDMmonteCarlos} 
\begin{figure}
 \centering
 \includegraphics[width=\hsize]{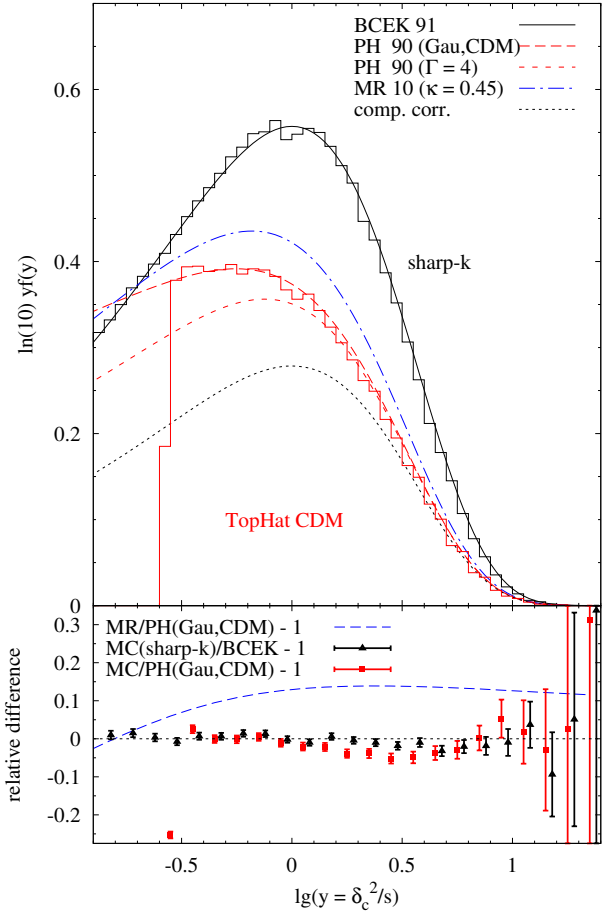}
 \caption{\emph{Top panel:} Distribution of the scale $s$ on which
   walks which first cross $\delta_c$ (histograms) for a $\Lambda$CDM 
   $P(k)$. The histogram which shows more objects at large $y =
   \delta_c^2/s$ is from walks with uncorrelated steps (labelled
   sharp-$k$); the other shows the result for walks with correlated
   steps due to TopHat smoothing. Solid curve shows the analytic
   prediction for uncorrelated steps from Bond \etal\ (1991); dotted
   curve is this divided by a factor of 2, for completely correlated
   steps; the two dashed curves show two implementations of the
   Peacock \& Heavens (1990) approximation for a Gaussian filter (see
   text for details); and the dot-dashed curve shows the approximation
   of Maggiore \& Riotto (2010a). \emph{Bottom panel:} Red squares
   show the residuals (with Poisson errors) between the walks with
   correlated steps and the scale-dependent implementation of the
   Peacock-Heavens approximation. Blue dashed curve shows the relative
   difference between the Maggiore-Riotto and Peacock-Heavens
   results. For comparison, the black triangles (which were given a
   small horizontal offset for clarity) show the residuals between the
   walks with uncorrelated steps and the corresponding analytic
   prediction of Bond \etal.}
 \label{vfvCDM}
\end{figure}

Figure~\ref{vfvCDM} shows the result of Monte-Carloing the first  
crossing distribution associated with a barrier of constant height 
$\delta_c$.  In the top panel,
the black histogram (the one which has more counts at 
large $y\equiv\delc^2/s$) shows the Monte-Carloed distribution for walks
with uncorrelated steps.  The solid curve going through it shows the 
associated analytic expression for the first crossing distribution 
(equation~\ref{fbcek} below).  
The agreement indicates that the numerical algorithm works.  

The red histogram shows the result when the steps are correlated.  
In practice, we transformed each walk with uncorrelated steps into 
one with correlations by applying smoothing filters of different 
scales following Bond \etal\ (1991).  In this case, the correlation 
depends on the form of the filter and on the shape of the initial 
linear theory power spectrum $P(k)$.  We used a Tophat smoothing filter 
and a $\Lambda$CDM power spectrum appropriate for 
 $(\Omega_m=0.25,\Omega_\Lambda=1-\Omega_m,h=0.7,\sigma_8=0.8)$.  
We then performed the same analysis as for the uncorrelated walks:  
find and store the scale on which $\delta_c$ is first crossed.  
Note that now the relation between $s$ and $M$ is modified compared 
to the previous case; we have checked that our algorithm does this 
correctly.  

The dotted curve shows the first crossing distribution associated with 
what we call completed correlated walks below (equation~\ref{fc} below), 
a limiting case that will prove useful for understanding many of the 
results to follow.  The short and long dashed curves show the 
Peacock-Heavens approximation (equation~\ref{fph} below), and the 
dot-dashed curve shows the approximation from equation 119 of Maggiore
\& Riotto (2010a). 
Both approximations have one free parameter, which we have set to the 
value appropriate for the walks shown in Figure~\ref{vfvCDM}.  
For Maggiore-Riotto, this parameter is $\kappa=0.45$.  
For Peacock-Heavens, the parameter $\Gamma$ is actually scale dependent 
(see Figure~\ref{GammaCDM} below):  the short dashed curve shows the 
result of setting $\Gamma=4$ and ignoring this dependence; the 
long-dashed curve, which provides  a better description of the
numerical solution, includes this dependence. (For reasons that we
discuss later, we calculated $\Gamma$ for the long-dashed curve
assuming a Gaussian rather than TopHat filter.)

The bottom panel of \fig{vfvCDM} shows the residuals (red squares)
with Poisson errors between the TopHat CDM walks and the
Peacock-Heavens approximation with scale-dependent $\Gamma$, 
and the relative difference between the Maggiore-Riotto result and the
Peacock-Heavens approximation (dashed blue curve). For comparison, we
also show the residuals between the walks with uncorrelated steps and
the corresponding analytic prediction from Bond \etal\ (1991) (black
triangles). The points for uncorrelated steps were given a small
horizontal offset for clarity.

This shows that the Peacock-Heavens approximation for the first
crossing distribution is more accurate than that of Maggiore \&
Riotto\footnote{The latter shows discrepancies at the level of
  $\sim15\%$, in keeping with the fact that their treatment is a
  linearization in $\kappa\sim0.4$.}. In what follows, we will show
that it is also more easily extended to treat more general power
spectra, more general smoothing filters, and more general barrier
crossing problems. (In private communications with us, Maggiore \&
Riotto have emphasized that they do not believe that equation 119 of
Maggiore \& Riotto 2010a can be applied directly, without further
computation, to the case of more general smoothing filters or power
spectra.) 

\begin{figure}
 \centering
 \includegraphics[width=\hsize]{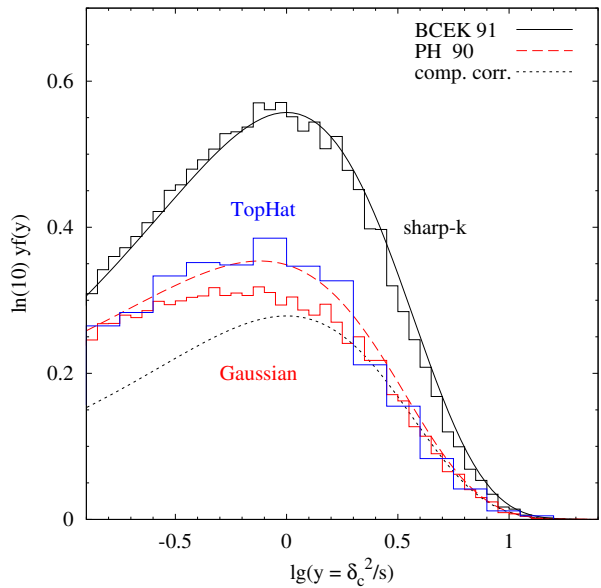}
 \caption{Distribution of the scale $s$ on which walks which first cross 
          $\delta_c$ (histograms) for $P(k)\propto k^{-1.2}$.   
          The histogram which shows more counts at large 
          $y = \delta_c^2/s$ is from walks with uncorrelated steps 
          (labelled sharp-$k$); 
          the other two histograms show the result for walks with 
          correlated steps, associated with TopHat and Gaussian smoothing 
          filters.  Solid curve shows the analytic prediction for 
          uncorrelated steps from Bond \etal\ (1991); dotted curve is 
          this divided by a factor of 2, for completely correlated steps; 
          and dashed curve shows the Peacock \& Heavens (1990) 
          approximation for a Gaussian filter, with $\Gamma=4.13$.} 
 \label{vfvConstant}
\end{figure}

\subsection{Power-law $P(k)$ and other smoothing filters}\label{monteCarlos}
Figure~\ref{vfvConstant} shows the first crossing distribution of a
constant barrier when $P(k)\propto k^{-1.2}$.  From top to bottom, 
the histograms show results for sharp-$k$ (black), TopHat (blue), 
and Gaussian (red) smoothing filters.  (We note again that the 
relation between $s$ and $M$ is filter dependent; our algorithm 
accounts for this correctly.)  The similarity of the TopHat and 
Gaussian histograms confirms a point made by Bond \etal\ (1991):  
When expressed as a function of $y$, the first crossing distribution 
associated with TopHat smoothing filters is approximately the same as 
for Gaussian filters.  (Since the relation between $y$ and $m$ is 
filter-dependent, this means that, when expressed as a function of $m$, 
the first crossing distribution does depend on filter.)  

The similarity of these distributions to those for the $\Lambda$CDM 
power spectrum illustrates another point that was implicit in the 
results of Bond \etal:  When expressed as a function of $y$, the 
first crossing distribution is approximately independent of 
power-spectrum.  (The choice $P(k)\propto k^{-1.2}$ is not special:  
we find qualitatively similar results when $P(k)\propto k^{-2}$.)  
Note that, for sharp-$k$ smoothing, this independence of $P(k)$ is 
exact.  

The solid and dotted curves are the same as in the previous Figure; 
they show the first crossing distributions associated with walks that 
have uncorrelated (sharp-$k$ filtering; equation~\ref{fbcek}) and 
completely correlated steps (equation~\ref{fc}).  
Notice that both the Gaussian/TopHat solutions approach the dotted 
curve asymptotically at large $y$, but lie above it at smaller $y$.  
We will have more to say about this later.  The dashed curve shows 
the Peacock-Heavens approximation (equation~\ref{fph} below) for 
Gaussian filtering (for which $\Gamma=4.13$).  It is in reasonable 
agreement with the Monte-Carlo solution, slightly overshooting at the 
peak.  It happens, just coincidentally, to provide a good description 
of the TopHat case.  This shows that the Peacock-Heavens approximation 
is rather accurate for a wide variety of power spectra and smoothing 
filters.  (We have checked that the relative differences between this 
approximation and that of Maggiore \& Riotto are approximately
independent of $P(k)$ and filter -- indicating that the latter may
have wider applicability, as it stands, than simply to TopHat
smoothing of a $\Lambda$CDM power spectrum.)

Having motivated why the Peacock-Heavens approximation is so 
interesting, we now discuss why it works, before using the insight 
gained to extend the approach to other barriers and barrier crossing 
problems.  

\subsection{Analytic approximation: Completely correlated steps}\label{nuwalks}
Walks with uncorrelated steps are jagged and stochastic.  
For such walks, the first crossing distribution is  
\begin{eqnarray}
 f_u(s)\,{\rm ds} &=& -\frac{\partial P_u(s)}{\partial s}  
    = -\frac{\partial\ }{\partial s}{\rm erf}(\delta_{\rm c}/\sqrt{2s})\\
    &=& \frac{\delta_c}{\sqrt{2\pi s}}\,e^{-\delta_c^2/2s}\,\frac{{\rm d}s}{s}
 \label{fbcek}
\end{eqnarray}
(Chandrasekhar 1943; Bond \etal\ 1991), where $P_u(s)$ is the
``survival probability'' that a randomly chosen walk has not crossed
the barrier $\delta_c$ prior to $s$.  

Walks with correlated steps are smoother.  However, before stating the
Peacock-Heavens approximation for walks with correlated steps, we
believe it is useful to study another case which can be solved
exactly and which is in some sense the opposite of the uncorrelated 
steps problem.  In this subsection,  we will be interested in the limit 
in which the walks are as smooth and deterministic as possible.  
In particular, we would like to think of such walks as having completely 
correlated steps, where by complete correlation we mean that the 
height of the walk at one time completely specifies its value at all 
other times.  

This notion of a completely correlated walk is, at first sight, 
somewhat ambiguous, since a walk which has the same height $\delta$ 
at all $s$ could be said to completely correlated, but a walk which 
is defined by a straight line from the origin through the point 
in question $(\delta,s)$, or indeed, any curve $f(s)$ whose value is 
completely specified by the pair $(\delta,s)$, would also be completely 
correlated.  
However, if we now require that the ensemble of such completely 
correlated walks also satisfy the constraint that the fraction of 
walks which lie above $\delta$ on scale $s$ equals 
 ${\rm erfc}(\delta/\sqrt{2s})/2$, i.e., obeys Gaussian statistics, 
then it must be that the number which specifies each member of this 
ensemble is $\nu = \delta/\sqrt{s}$.  Thus, a $\nu=1$ walk is one 
which has height $\delta=\sqrt{s}$ on scale $s$; a $\nu=2$ walk is 
one which has height $2\sqrt{s}$ on scale $s$, etc.  Notice that 
these are walks which are {\em not} constant height $\delta$, but 
constant $\nu$:  their height scales as $\sqrt{s}$.  We will refer 
to this family of walks as having completely correlated steps.  

Now consider a barrier of constant height $\delta_{\rm c}$.  
The first crossing distribution associated with this family of walks 
will depend on the distribution of $\nu$.  If this distribution is 
Gaussian, $p(\nu) = \exp(-\nu^2/2)/\sqrt{2\pi}$, then the 
corresponding survival probability is
\begin{equation}
 P_{\rm c}(s) = \frac{1}{2}\left (1+{\rm erf}(\delta_{\rm c}/\sqrt{2s})
 \right). 
 \label{Fdeterministic}
\end{equation}
As a result, 
\begin{equation}
 sf_c(s) = -\frac{\partial P_{\rm c}(s)}{\partial\ln s} =
 -\frac12 \frac{\partial P_u(s)}{\partial\ln s} =
 \frac{sf_u(s)}{2};  
 \label{fc}
\end{equation}
this differs from the case of completely uncorrelated steps by the 
factor of two in the denominator.  The origin of this factor is clear:  
if $\delta_{\rm c}>0$, then walks having $\nu<0$ can never cross 
$\delta_{\rm c}$.  For a Gaussian distribution of $\nu$, this means 
half the walks never cross $\delta_{\rm c}$.  This is a novel way to 
understand just what it is that Press \& Schechter (1974) derived:   
their expression describes the first crossing distribution of walks 
having completely correlated steps (in the sense described above).  

\begin{figure}
 \centering
 \includegraphics[width=\hsize]{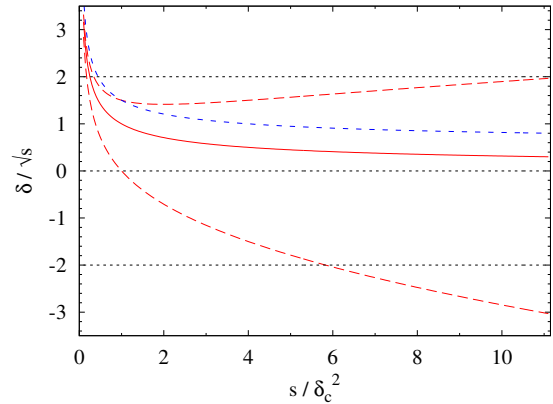}
 \caption{Completely correlated walks (horizontal lines) in the
   presence of various barriers. The solid curve is a constant barrier
   $B(s)=\delc$. The long-dashed red curves are for linear
   barriers $B(s)=\delc(1+\beta s/\delc^2)$, with $\beta=-1$ (lower)
   and $\beta=0.5$ (upper), the latter displaying a minimum at
   $s=S_{\rm crit}=\delc^2/\beta$. The short-dashed blue curve is the
   square-root barrier $B(s)=\delc(1+\beta\sqrt{s}/\delc)$, which
   mimics the constant barrier in that $B(s)/\sqrt{s}$ has a minimum
   at $S_{\rm crit}\to\infty$. See text for a discussion.} 
 \label{ccwalks}
\end{figure}

Before moving on, note that it is trivial to extend this analysis 
to `moving' barriers, whose height depends monotonically on $s$.  
For barriers $B(s)$ which decrease with $s$, one simply replaces 
$\delta_{\rm c}\to B(s)$ in the expression above,
\begin{align}
 P_{\rm c}(s) &= \frac{1}{2}\left (1+{\rm erf}(B(s)/\sqrt{2s})
 \right)\,, \label{PcdecreasingB}\\
 sf_c(s) &= -\frac{\partial P_{\rm c}(s)}{\partial\ln s}\nonumber\\
&= -\frac{\partial\ln(B/\sqrt{s})}{\partial\ln s}
 \frac{B(s)}{\sqrt{2\pi s}}\,e^{-B(s)^2/2s}\,. 
\label{fcdecreasingB}
\end{align}
For barriers which increase with $s$, this is slightly more involved,
as shown by \fig{ccwalks} which plots $\delta/\sqrt{s}$
vs. $s$. 

Completely correlated walks correspond to horizontal lines on 
this plot. It is then obvious that for barriers $B(s)$ such that
$B(s)/\sqrt{s}$ has a minimum at some $s=S_{\rm crit}$, there is no
crossing of the barrier for $s>S_{\rm crit}$. (Since $B(s)$ is
monotonically increasing, $B(s)/\sqrt{s}$ monotonically
increases for $s>S_{\rm crit}$.) As a result, the survival probability
becomes 
\begin{equation}
  P_{\rm c}(s) =
   \begin{cases} 
     \frac12 \left(1+{\rm erf}[B(s)/\sqrt{2s}]\right)\,, & {\rm if}\quad s\le S_{\rm crit}\\
     \frac12\left(1+{\rm erf}[B(S_{\rm crit})/\sqrt{2S_{\rm crit}}]\right)\,,
     & {\rm if}\quad s > S_{\rm crit}\,
   \end{cases}
 \label{increasingB}
\end{equation}
and the rate $f_c(s)$ is given by \eqn{fcdecreasingB} until $s=S_{\rm
  crit}$, at which point it becomes and stays zero. We will return to
this point later.

To end this subsection, we note that an interesting example of an
increasing barrier is one which increases as the square-root of $s$:   
\begin{equation}
 B(s) = \delta_c (1 + \beta \sqrt{s}/\delta_c).
\end{equation}
In this case, although completely correlated walks of height
$\nu<\beta$ will never cross $B(s)$, we have $S_{\rm crit}\to\infty$,
so the complication associated with $s>S_{\rm crit}$ does not
arise. This barrier is a convenient approximation to the `moving 
barrier' associated with halos which form from an ellipsoidal collapse 
(Moreno \etal\ 2009). 
While showing nontrivial $s$ dependence, it retains the simplicity 
of the constant barrier (see \fig{ccwalks}). The rate of increase of 
this barrier ($\sim s^{1/2}$) also serves as the dividing line beyond which
$S_{\rm crit}$ takes finite values and the complexity of 
\eqn{increasingB} comes into play.  
 Shen \etal\ (2006) argue that this form, with
$\beta<0$, is also a convenient approximation to the barrier
associated with the formation of sheets. 

\subsection{Correlated, but not deterministic, steps}\label{cwalks}
The two limiting cases -- of maximally stochastic and
completely deterministic walks -- serve as useful guides for the 
construction of the first crossing distribution when there is some, 
but not complete, correlation between steps.  E.g., for the case 
of very weak or very strong correlations, one might imagine perturbing 
around one or another of these solutions.  We will argue below that 
the Peacock-Heavens approximation may be thought of as perturbing 
around the case of complete correlation.  

We first briefly restate their approximation, following the 
presentation of it in Bond \etal\ (1991).  The Peacock-Heavens
approximation derives from noting that the height of a walk on scale
$s$ is correlated with its height on scales that are within $s\Gamma$
of it, where $\Gamma$ is a parameter that depends on details of the
filter, and will be defined below. One then asserts that the walk can
be broken up into independent segments of length $s\Gamma$, and
requires that the height of the walk is below the barrier after each
step.  It may be helpful to think of this approximation as stating
that, of the fraction of walks that are below $\delta_c$ after $n$
steps, one must take the fraction that were also below after $n-1$
steps, and the fraction of these that were below after $n-2$ steps and
so on.  If we define
\begin{equation}
 c(<\delta|s) \equiv \int_{-\infty}^{\delta/\sqrt{s}} {\rm d}x\, 
                        \frac{\exp(-x^2/2)}{\sqrt{2\pi}},
\label{cdef}
\end{equation}
then the Peacock-Heavens ansatz for the 
survival probability in the presence
of a single barrier of height $\delta_{\rm c}$ is 
\begin{equation}
 P_{\rm PH}(s_n) = c(<\delta_{\rm c}|s_n) \prod_{i=1}^{n-1}
 c(<\delta_{\rm c}|s_i),
\label{PHsurvdiscrete}
\end{equation}
where the spacing between the $s_i$ is chosen such that each step is 
approximately independent of the previous ones.  
One then expresses the product as the exponential of a sum over logs, 
and then replaces the sum by an integral.  
If we define 
\begin{equation}
 p(s) = c(<\delta_c|s)
 \label{PHconstant}
\end{equation}
and
\begin{equation}
 E_{\rm c}(s) \equiv \exp\left(\int_0^s \frac{{\rm d}s^\prime}{\Gamma
   s^\prime}\, \ln p(s^\prime)\right),
 \label{Ecorrection}
\end{equation} 
then 
\begin{eqnarray}
 sf_{\rm PH}(s) &=& -\frac{\partial P_{\rm PH}(s)}{\partial \ln s} 
         = -\frac{\partial\,\left[p(s)\,E_{\rm c}(s)\right]}{\partial \ln s}
             \nonumber\\
   &=& E_{\rm c}(s)\left\{sf_c(s) - \frac{p(s)\,\ln p(s)}{\Gamma} \right\},
 \label{fph}
\end{eqnarray}
where $f_c$ was defined in equation~(\ref{fc}).  
The Appendix discusses why this is only an approximation to the exact 
solution.  

What remains is to determine $\Gamma$.  
But before doing so, note that as $\Gamma\to\infty$, then the term in 
the exponential of equation~(\ref{Ecorrection}) $\to 0$, 
so $E_{\rm c}$ itself $\to 1$, and in equation~(\ref{fph}) 
$f_{\rm PH}(s)\to f_c(s)$, 
the distribution for completely correlated walks.  
Hence, if we view $E_{\rm c}(S)$ as a series in $1/\Gamma$, then 
one may think of the Peacock-Heavens approximation as perturbing 
around the case of smooth completely correlated walks.  In fact, we
can rewrite the survival probability as
\begin{equation}
 P_{\rm PH}(s) = P_{\rm c}(s)\,E_{\rm c}(s)\,,
 \label{Pph}
\end{equation}
where $P_{\rm c}(s)$ was defined in equation~(\ref{Fdeterministic}), and
think of $E_{\rm c}(s)$ as a correction to the completely correlated case.
On the other hand, note that the $\Gamma\to 0$ limit does {\em not} 
reduce to the expression for uncorrelated steps, and is in fact not
even well-defined.  

\begin{figure}
 \centering
 \includegraphics[width=\hsize]{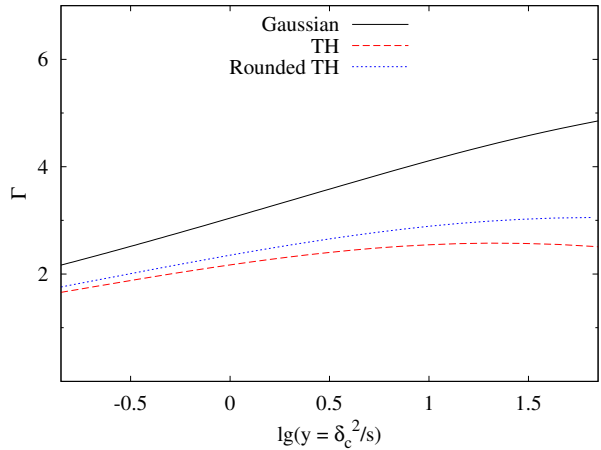}
 \caption{The Peacock-Heavens correlation length $\Gamma$ for a CDM
   power spectrum, as a function of variance $s$, with three different
   filters : Gaussian (solid black) with $W_{\rm G}(x)=e^{-x^2/2}$
   where $x=k R_{\rm G}$; TopHat (dashed red) with $W_{\rm
     TH}(x)=(3/x^3)[\sin(x)-x\cos(x)]$ where $x=k R_{\rm TH}$; and a
   rounded TopHat (dotted blue) for which $W_{\rm rTH}(x) = W_{\rm
     TH}(x)e^{-(1/2)(x/10)^2}$ where $x=k R_{\rm rTH}$. A given value
   of $s$ will, in general, correspond to different values of $R_{\rm
     G}$, $R_{\rm TH}$ and $R_{\rm rTH}$.}
 \label{GammaCDM}
\end{figure}

For Gaussian smoothing filters and generic power spectra,
the Peacock-Heavens
prescription for $\Gamma$ is
$\Gamma = 2\pi\ln(2) [(\sig_0^2\sig_2^2)/\sig_1^4-1]^{-1/2}$, 
where the $\sig^2_j$ were defined in \eqn{sig2jdef}. 
This can be written as
\begin{equation}
 \Gamma = 2\pi\ln(2)\,\sqrt{\gamma^2/(1-\gamma^2)}\,.
\label{Gamdef}
\end{equation}
To see what this implies, suppose that $P(k)\propto k^n$.  
Then $\gamma^2 = (3+n)/(5+n)$, making $\Gamma = 2\pi\ln(2)
\sqrt{(3+n)/2}$. For $n=(-1.2,-2)$, $\Gamma = (4.13,3.08)$.  For a more
realistic CDM spectrum, \fig{GammaCDM} shows that $\Gamma$ varies with
scale (approximately logarithmically with $s$), with similar numerical 
values. But if the filter is a TopHat in real space, then some of the integrals 
appearing in \eqn{Gamdef} may diverge, so one must compute
$\Gamma$ slightly differently.  In Appendix B, we show that for
power law spectra $\Gamma = 2\pi\ln(2)\sqrt{(n+1)(n+3)/(n-3)}$, so the
Peacock-Heavens procedure is well-defined only if $-3<n<-1$. Note that
the TopHat filter is known to be analytically troublesome with power
law spectra, with even the variance $s$ being undefined for
$n\geq1$. The TopHat is well-behaved with the CDM spectrum though, and
\fig{GammaCDM} shows that $\Gamma$ is well-defined in this
case. Nevertheless, as \figs{vfvCDM} and~\ref{vfvConstant} show, the
numerical results for TopHat filtering are actually better described
by the analytical approximation for Gaussian filtering.

The oscillatory behaviour of the Fourier transform of the TopHat
filter, results in TopHat smoothed walks being less correlated than
Gaussian smoothed walks. One therefore expects that the value of
$\Gamma$ for the TopHat filter should be smaller than that for the
Gaussian. Numerically, since the first crossing distributions for the
two filters are not very different, the difference in values of
$\Gamma$ should at most be a factor of order unity, and this is true
for the CDM spectrum. The dramatic discrepancy for power 
law spectra is most likely caused by the known bad behaviour of the
TopHat in this case, combined with the fact that the Peacock-Heavens 
procedure is a perturbation around the completely correlated case, and
therefore performs better for Gaussian smoothed walks, which are
closer to complete correlation than are TopHat walks. 

One possibility for alleviating the problem of the TopHat filter with
power law spectra is to round the sharp edge of the TopHat
(of scale $R_{\rm TH}$) with, say, a Gaussian filter of a smaller scale 
$\ep R_{\rm TH}$ (Bond \etal\ 1991).  The resulting $\Gamma$ is 
well-defined for all $n>-3$, and smaller than the Gaussian. 
\fig{GammaCDM} shows the result for a CDM spectrum, using $\ep=0.1$. 
In the remainder of the paper, we will use the Peacock-Heavens analysis 
for Gaussian smoothing and power law spectra, since this combination 
is simple, well-defined and accurate. The discussion above (and a 
comparison of Figures~\ref{vfvCDM} and~\ref{vfvConstant}) shows that, 
with some care, our conclusions can be generalized to more realistic 
power spectra and filters.  

\section{Moving barriers}\label{moving}
One of the virtues of the Peacock-Heavens approximation is that it 
is easy to see how equation~(\ref{Pph}) should be extended to the case
when $\delta_c$ is no longer constant.  The logic behind writing the
correction factor $E_{\rm c}(s)$ as in equation~(\ref{Ecorrection})
does not depend on the barrier being constant. Therefore, for
\emph{any} moving barrier $B(s)$, this suggests setting 
\begin{equation}
 p(s) = c(<B(s)|\sqrt{s})\,,
 \label{PHmoving}
\end{equation}
in equation~(\ref{Ecorrection}). 

\begin{figure}
 \centering
 \includegraphics[width=\hsize]{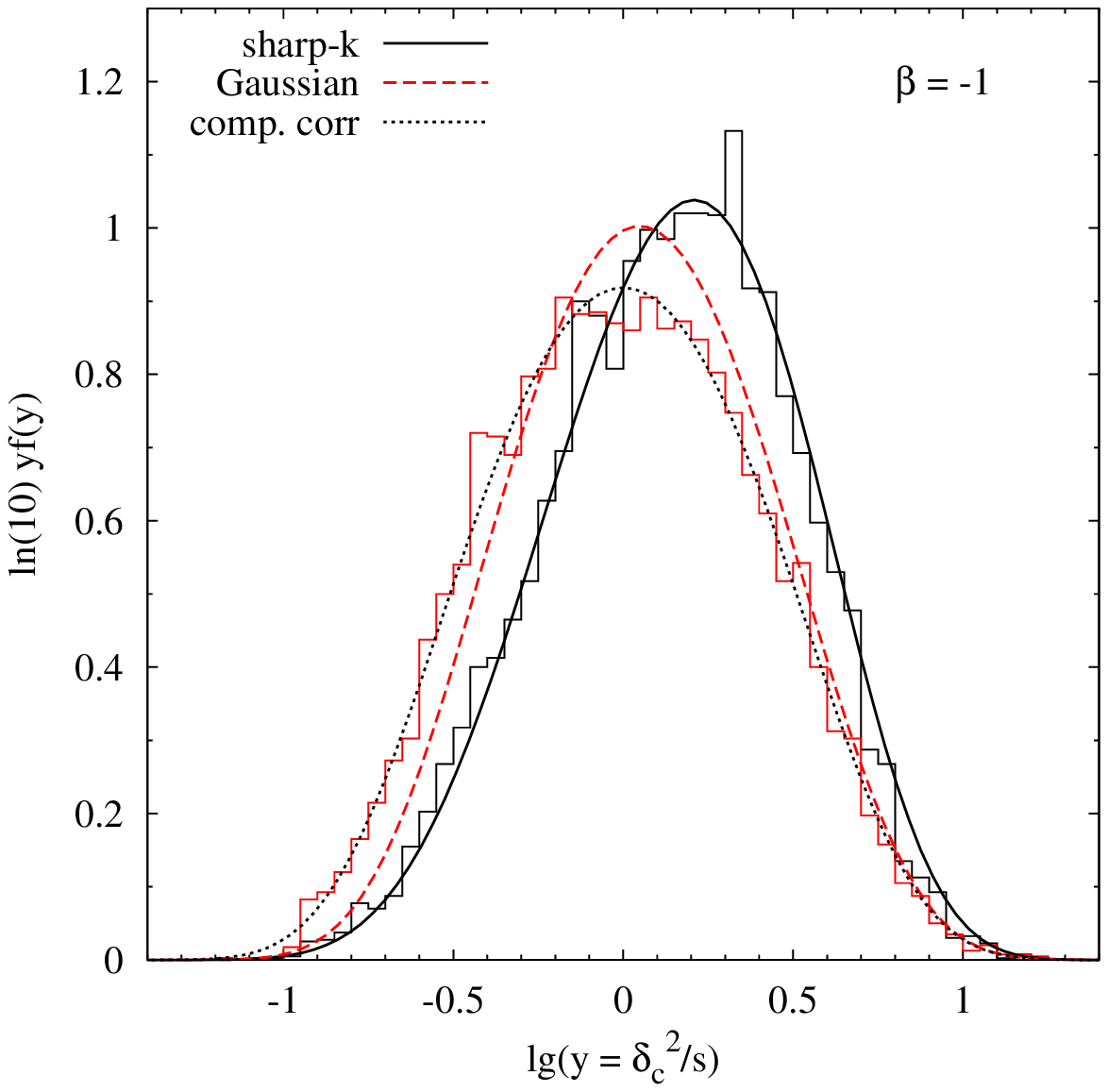}
 \includegraphics[width=\hsize]{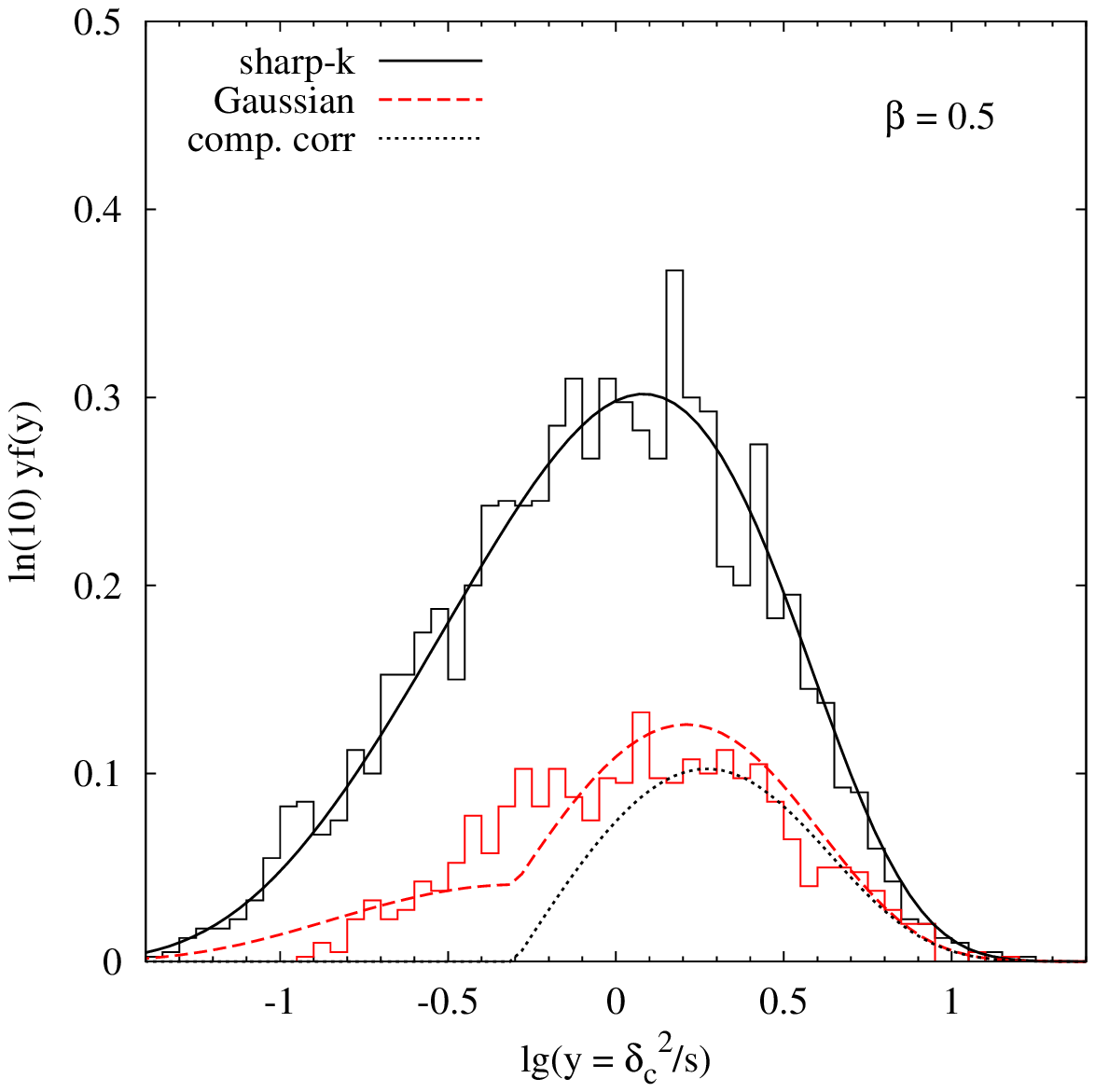}
 \caption{Distribution of the scale $s$ on which walks which first cross 
          the barrier $\delta_c(1 + \beta s/\delta_c^2)$ (histograms).   
          The histogram which shows more 
          objects at large $y = \delta_c^2/s$ is from walks with 
          uncorrelated steps; the other histogram shows the result 
          for walks with correlated steps.  Solid curve shows the 
          analytic prediction for uncorrelated steps from 
          Sheth (1998); dashed curve shows our extension of the 
          Peacock \& Heavens (1990) approximation for correlated
          steps. Dotted curve shows the expression for completely
          correlated walks.}
 \label{vfvLinear}
\end{figure}

On the other hand, the discussion in Section~\ref{nuwalks} shows that
the survival probability for completely correlated walks
 $P_{\rm c}(s)$ must be handled with care, depending on the nature of the
moving barrier. For barriers $B(s)$ which are decreasing functions of
$s$, we can still make the simple replacement $\delta_c\to B(s)$ in
equation~(\ref{Fdeterministic}). Figure~\ref{vfvLinear} shows that,
for barriers of the form  
\begin{equation}
 B(s) = \delta_c (1 + \beta s/\delta_c^2)
 \label{linearB}
\end{equation}
with $\beta<0$, the resulting expression for the first crossing
distribution (dashed curve) describes the Monte-Carlo solution
rather well.  We also note that, for this case, the expression for
\emph{completely} correlated walks \eqn{fcdecreasingB} (dotted curve) 
actually describes the numerical solution more accurately.  We return 
to this in the final Discussion section.

For barriers which increase with $s$, such that $B(s)/\sqrt{s}$ has a
minimum at some $s=S_{\rm crit}$, Section~\ref{nuwalks} shows
that we must use equation~(\ref{increasingB}) rather
than~(\ref{Fdeterministic}) to calculate the completely correlated
survival probability $P_{\rm c}(s)$ to use in
equation~(\ref{Pph}). This correctly accounts for walks that never
cross the barrier. Figure~\ref{vfvLinear} shows that this prescription
works well in describing the Monte-Carlo solution for a linearly
increasing barrier (\ref{linearB}) with $\beta>0$, although there is a
discrepancy at small $y$. Notice that in this case, the completely
correlated expression does not perform well.

\section{Constrained walks}\label{constrained}
The Peacock-Heavens approximation can also be extended to describe the 
first crossing distribution of walks which are conditioned to pass 
through some non-zero $(\delta,S)$.  In many respects, this problem 
highlights the difference between walks with uncorrelated steps, and 
those which are completely correlated.  

Since a completely correlated walk is specified by a single number, 
if it is known to have height $\delta_S$ on scale $S$, then it will 
have height $\delta_S\,\sqrt{s/S}$ on scale $s$.  I.e., the expected 
distribution of heights on scale $s$ is a delta function, centered on 
$\delta_S\,\sqrt{s/S}$.  On the other hand, for uncorrelated walks, 
this distribution is Gaussian with mean $\delta_S$ and variance $s-S$ 
(we have assumed $s>S$).  This means that, except for a shift of origin, 
the conditional walk follows the same statistics as the unconditioned 
one.  Namely, one simply sets
 $\delta_{\rm c}\to \delta_{\rm c}-\delta_S$  and $s\to s-S$ in 
equation~(\ref{fbcek}) for the first crossing distribution.  

Notice that the determinstic walk increases its height by an amount
$\delta_S (\sqrt{s/S}-1)$ whereas stochastic walks have no change 
on average, but some walks will reach large heights because the rms 
increases as $\sqrt{s-S}$.  
The general case will lie somewhere in between these two extremes:  
we might generically expect a milder increase in the expected height 
compared to the deterministic case, with a narrower distribution around 
the mean compared to the stochastic case.  

\subsection{A scaling ansatz based on bivariate statistics}
To extend the Peacock-Heavens approach to describe constrained walks, 
we will make an ansatz which we justify later.  Our ansatz is that the 
conditional distribution is just a rescaled version of the unconditional 
one, where the scaling variable is inspired by bivariate Gaussian 
statistics.  Namely, we define 
\begin{equation}
 \nu_{10} \equiv \frac{\delta_{\rm c} - r\, \delta\,\sqrt{s/S}}{\sqrt{s(1 - r^2)}}
          \equiv \frac{\delta_{\rm c} - (S_\times/S)\, \delta}
                      {\sqrt{s - (S_\times/S)^2S}},
 \label{v10}
\end{equation}
where $r \equiv S_\times/\sqrt{sS}$ 
and 
\begin{equation}
 S_\times \equiv 
  \int \frac{{\rm d}k}{k}\,\frac{k^3\, P(k)}{2\pi^2}\, W(kR_s)W(kR_S).
\end{equation}
Notice that the integral above is similar to that which defines $s$ and 
$S$, the only difference being that here the two smoothing filters have 
different scales.  Our ansatz is that, when expressed as 
$\nu_{10} f(\nu_{10})$, the conditional first crossing distribution will 
have the same shape as the unconditional distribution $\nu f(\nu)$, a 
point we will return to shortly.

Equation~(\ref{v10}) accounts for the fact that if the field is 
constrained to have value $\delta_S$ on scale $S$, then its value on 
scale $s$ will be distributed around a mean value of 
$(S_\times/S)\,\delta_S$.  For uncorrelated steps $S_\times = S$, so 
$s(1-r^2) = s - S$ and hence 
$\nu_{10} = (\delta_{\rm c} - \delta_S)/\sqrt{s-S}$.  
This corresponds to simply shifting the origin of the walk from 
$(0,0)$ to $(\delta_S,S)$, as expected from the previous discussion.  
Completely correlated walks have $r=1$; in this limit, our expression 
for $\nu_{10}$ correctly indicates no scatter around a mean value of 
$\delta_S\sqrt{s/S}$.  

For Gaussian filtering of a power-law spectrum 
   $r = [2R_SR_s/(R_S^2+R_s^2)]^{(n+3)/2}$ 
and $S\propto R_S^{-(n+3)}$,
making 
$r \sqrt{s/S} = [2/(1 + (S/s)^{2/(3+n)})]^{(n+3)/2}$, 
and $s(1-r^2) = s - S [2/(1 + (S/s)^{2/(3+n)}]^{(n+3)}$.
To get some intuition into what this implies, it is helpful to 
consider the limit $R_s\ll R_S$.  In this case, $S\ll s$, so 
$\nu_{10} \approx (\delta_{\rm c} - 2^{(n+3)/2}\delta_S)/\sqrt{s -
  2^{3+n}S}$. This corresponds to a shift of origin that is larger
than for the uncorrelated case, but much smaller than for the
completely correlated case, and a variance that is slightly smaller
than for the uncorrelated case, but much larger than for the 
completely correlated case.  The discussion above applies for other
smoothing filters too,  except that the numerical coefficient
$2^{(n+3)/2}$ will change. In Appendix B we give some details for the 
TopHat filter.

\begin{figure}
 \centering
 \includegraphics[width=\hsize]{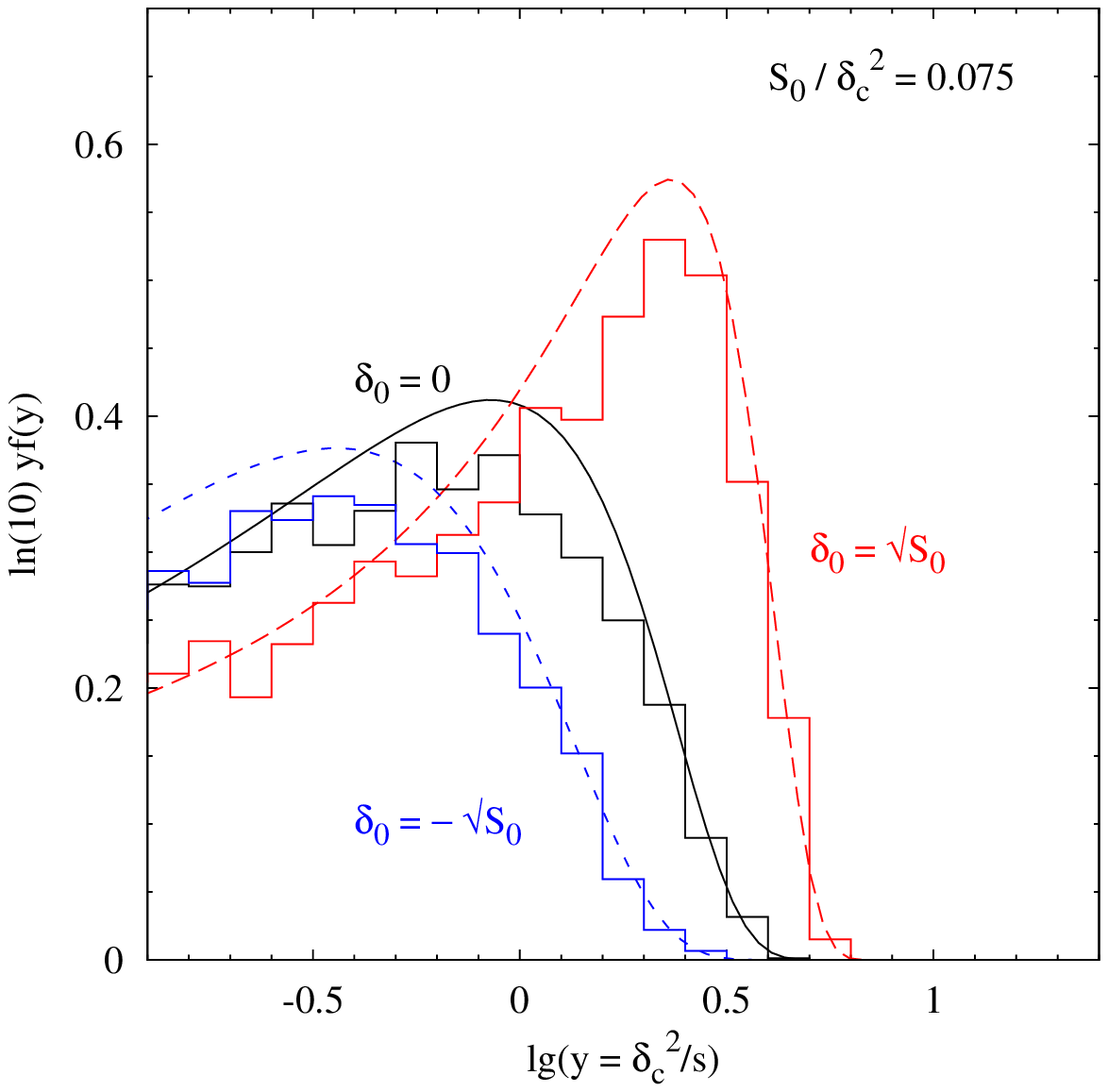}
 \includegraphics[width=\hsize]{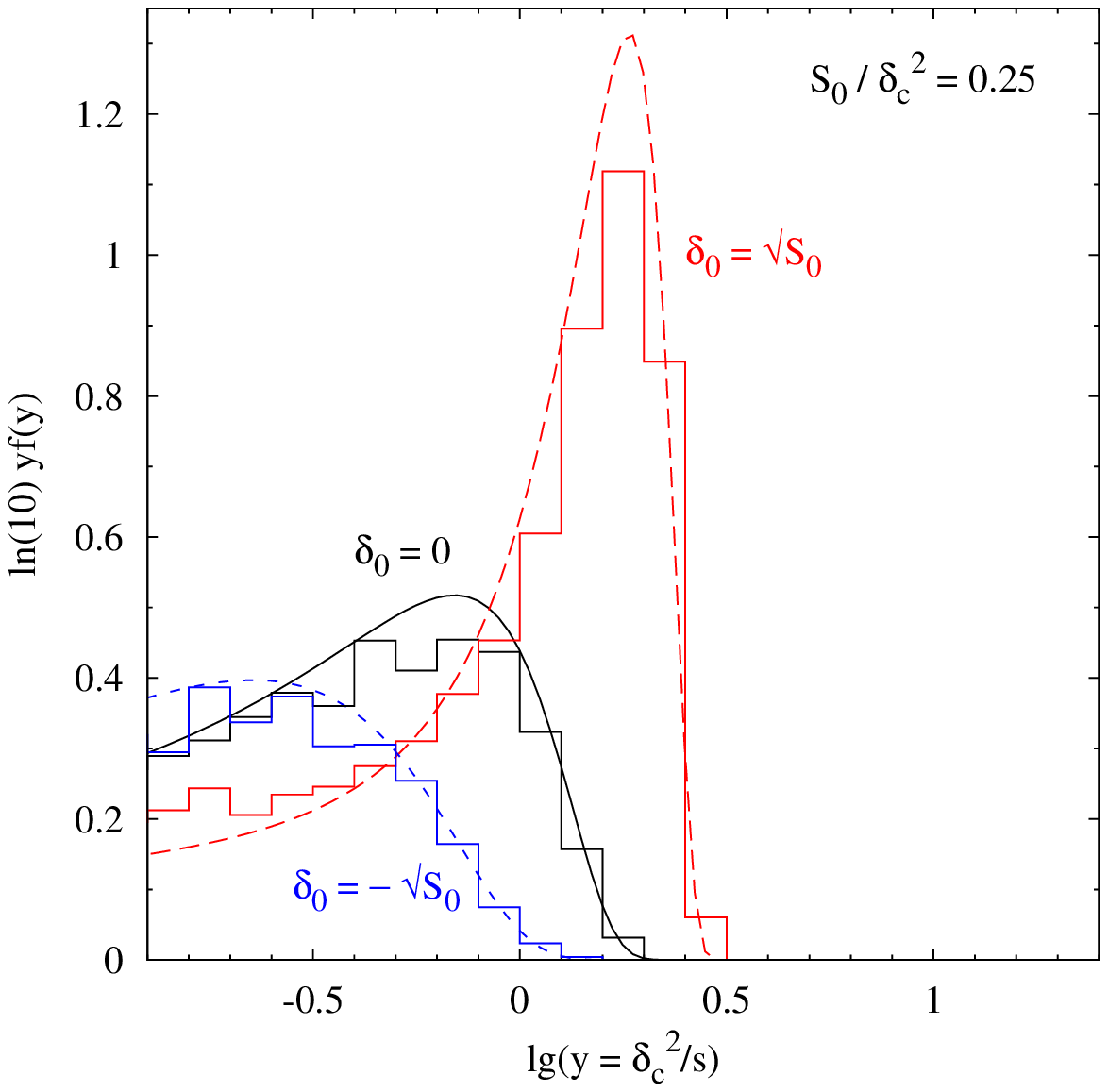}
 \caption{First crossing distribution of a barrier of height 
          $\delta_{\rm c}$ by the subset of walks which 
          are conditioned to pass through $(\delta_0,S_0)$, for 
          a few choices of $\delta_0$ (as labelled). 
          Short dashed, solid and long-dashed curves show the 
          analytic prediction from our extension of the 
          Peacock \& Heavens (1990) approximation (equation~\ref{cCond}
          in equation~\ref{fph}), 
          for Gaussian smoothing of a Gaussian field with 
          $P(k)\propto k^{-1.2}$.}
 \label{vfvCond}
\end{figure}

\subsection{Comparison with Monte-Carlos}
For the Peacock-Heavens approximation, our ansatz means that we set 
\begin{equation}
 c(<\delta_{\rm c}|s) \to c(<\delta_{\rm c},s|\delta,S) 
                     = \frac{1}{2}\left[1 + {\rm
                       erf}(\nu_{10}/\sqrt{2})\right] 
 \label{cCond}
\end{equation}
in equation~(\ref{PHconstant}) when computing 
equations~(\ref{Ecorrection}), (\ref{Pph}) and~(\ref{fph}).  
Figure~\ref{vfvCond} shows that this works reasonably well.  
The histograms show the first crossing distribution of a barrier of 
height $\delta_{\rm c}$ by the subset of walks which are conditioned 
to pass through $(\delta_0,S_0)$, for a few choices of $\delta_0$, 
and the smooth curves show our extension of the Peacock-Heavens 
approximation.
Whereas the qualitative trend is easy to understand -- walks which start 
closer to the barrier (i.e. large $\delta_0$) cross it after fewer 
steps so there are few left to first cross at $s\gg S$ -- our approach 
does a reasonable job of quantifying the effect.  

\begin{figure}
 \centering
 \includegraphics[width=\hsize]{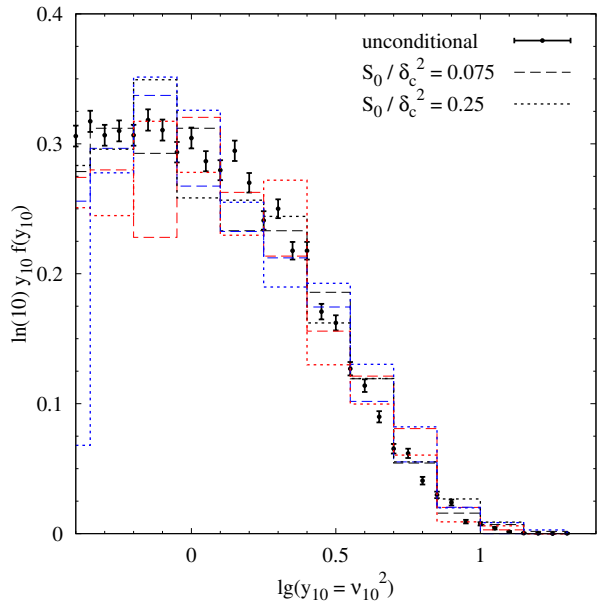}
 \caption{Conditional first crossing distributions of the previous 
          Figure, expressed as a function of the scaling variable 
          $y_{10}\equiv \nu_{10}^2$ (equation~\ref{v10}).  Symbols 
          with error bars show the unconditional distribution 
          from Figure~\ref{vfvConstant}. }
 \label{y10fy10}
\end{figure}

Our ansatz is that the conditional distribution is just a rescaled 
version of the unconditional one.  We have tested it more directly 
as follows.  Figure~\ref{y10fy10} shows the result of transforming 
each of the first crossing values $s$ into the associated $\nu_{10}$, 
and then plotting the conditional first crossing distribution as 
$y_{10}f(y_{10})$, where $y_{10}\equiv \nu_{10}^2$.    
If our ansatz is good, then all the curves of the previous Figure 
should define a single universal curve.  At least over the range of 
$y_{10}$ we have shown ($y_{10}>0.3$), and for the range of $S_0$ we 
have studied ($S_0/\delta_{\rm c^2}<0.3$ or so) they do.  Moreover, 
this universal curve has the {\em same} shape as that for unconditional 
walks -- which we have shown using symbols with error bars -- indicating 
that our ansatz is indeed a good one.  In particular, this indicates 
that most (if not all) of the discrepancy between the smooth curves and 
Monte-Carlo'd histograms in Figure~\ref{vfvCond} is due to the 
inaccuracy of the Peacock-Heavens approximation for unconditioned walks.  

Our ansatz, which allows one to transform any unconditional 
distribution into a conditional one (the accuracy of the resulting 
curve will be limited by that of the unconditional distribution, of 
course), allows us to provide some insight into the recent work of 
De Simone et al. (2011b), who present a path-integral analysis of the 
constrained walk problem.  
Their equation~(B30)\footnote{In an earlier
  version, their expressions B29-B31 contained an error, as can be
  seen by requiring consistency with Ma \etal\ 2011, or by simply 
  demanding that the limit $S_m\to0$, $\delta_0\to0$ reproduce the
  Maggiore-Riotto unconditional distribution. The factor
  $\delta_c-\delta_b$ should be $\delta_b$ in the last line of each of
  equations B29-B31. The error was corrected in a subsequent version
  of the paper, after it was pointed out by us.} 
with $S_m\to0$ (derived using the path integral
machinery after linearizing in $\kappa$) is, in fact, equivalent to
simply replacing  $\nu\to [\delta_c-\delta_0(1+\kappa)]/\sqrt{s}$ in 
their expression for the unconditional mass function, and then only 
keeping terms up to linear order in $\kappa$. 
Since their $\kappa$ is the $S\to 0$ limit of our $S_\times/S-1$, 
we conclude that our simple ansatz of $\nu\to\nu_{10}$ reproduces the 
detailed path integral result in the limit where path integral
calculations have been performed (i.e. to linear order in $\kappa$).  
Moreover, when $S$ is small but not zero, Figures~\ref{vfvCond} 
and~\ref{y10fy10} show that the same ansatz correctly describes the 
numerical solution, without having to linearize in $S_\times/S-1$ 
(which, in any case, is not small compared to unity).  
Of course, in our ansatz, $S_\times/S$ depends on scale, and appears 
in the denominator of $\nu_{10}$ as well. 

\subsection{Extensions and generalizations}
Our treatment of the first crossing of \delc\ by walks 
which are constrained to pass through some $(\delta,S)$ is easily 
extended in two ways.  One is to the `two-barrier' problem:  
given that the walk first crossed the barrier $\delta_{\rm c0}$ on 
scale $S_0$, what is the probability of first crossing the barrier 
$\delta_{\rm c1}>\delta_{\rm c0}$ on scale $S_1>S_0$?  In this case,
one uses
 $c(<\delta_{\rm c}|s) \to c(<\delta_{\rm c1},s|\delta_{\rm c0},S_0)$ 
for $S_0<s<S_1$, but $c(<\delta_{\rm c}|s) \to
c(<\delta_{\rm c0},s|\delta_{\rm c0},S_0)$ for $s<S_0$. It will be
interesting to test how well this simple extension does, compared to 
a numerical solution. In principle, we could go on to estimate merger
rates from the limit in which $\delta_{\rm c0}\to \delta_{\rm c1}$
(following Lacey \& Cole 1993).  In this limit, however, most walks
will cross the barrier within a few steps.  As a result, most
crossings will happen before the walk has travelled a distance that is
of order the correlation length $\Gamma S$, so the Peacock-Heavens
approximation is no longer expected to work.

The second generalization is the introduction of more constraints on 
the walks before (or after) they cross \delc.  Again, to 
incorporate these, one must simply modify equations~(\ref{cCond}) 
and~(\ref{v10}) appropriately, since the net effect of these additional 
constraints will be to change the mean and variance of the typical walk 
height on scales that are different from the constrained scales.

\section{Discussion}
We have presented an analysis of the first crossing distribution of 
a moving barrier by random walks having correlated steps -- the first 
analysis to explicitly compare analytic approximations with numerical 
(Monte-Carlo) solutions.  
For walks with uncorrelated steps, exact solutions are only known 
for a handful of special cases, although a good analytic approximation 
is available for the general case (Lam \& Sheth 2009).   
However, in the limit in which steps are completely correlated, so 
the walks are smooth and deterministic rather than jagged and 
stochastic, we showed that this solution is straightforward, at least 
for barriers which are monotonic functions of time/scale 
(Section~\ref{nuwalks}).  

For the more general case of some, but not complete, correlation 
between steps, Peacock \& Heavens (1990) provide a simple
approximation for the first crossing distribution of a barrier of
constant height (equation~\ref{fph}).  We showed that their
approximation can be thought of as a correction to the solution for
completely correlated steps (Section~\ref{cwalks}).  
The correction involves a suitably defined correlation length
$\Gamma$ (equation~\ref{Gamdef}),  
whose inverse can actually be thought of as an expansion parameter;
complete correlation is the limit $1/\Gamma\to 0$. 
While the prescription itself does not involve a perturbation in
$1/\Gamma$, one can understand that the approach works well 
(Figure~\ref{vfvConstant}) because  $1/\Gamma\approx 0.2$
(Figure~\ref{GammaCDM}).  We argued that this is also why the approach
works better for Gaussian than for TopHat smoothed walks -- the
latter have smaller values of  $\Gamma$.

We also showed (Figure~\ref{vfvCDM}) that, 
when using a constant barrier height without scatter,
the older Peacock-Heavens approximation was more accurate than that of
Maggiore \& Riotto (2010a).   
This more recent work, based on field-theoretic methods, explicitly 
perturbs around the solution for completely uncorrelated steps.  
In this case, the perturbation parameter $\kappa$, which should be 
$\ll 1$, is actually of order $1/2$.  Since $\kappa$ is larger than one 
would like, and $\Gamma$ is smaller, it is certainly interesting to 
attack the problem of walks with correlated steps from both
directions. (See, e.g., Corasaniti \& Achitouv 2011a for a recent
extension of the path integral approach to moving barriers.)
Additionally, as we discuss below and in the Appendix, it is also
interesting to explore the accuracy of these approaches when extended
to models in which the barrier height shows a scale-dependent scatter
rather than being deterministic.

In Section~\ref{moving}, we showed how to extend the Peacock-Heavens 
approximation to handle moving barriers.  For barriers which 
decrease with time/scale, this extension is remarkably simple and 
remarkably accurate (Figure~\ref{vfvLinear}).  
For barriers which increase sufficiently steeply with time/scale, 
the extension is slightly more complex, but still quite accurate.  
We summarize our results here: for a barrier
$B(s)$ which is monotonic in $s$, the first crossing distribution is
the derivative of the survival probability $f(s) = -\p_sP(s)$, where
\be
P(s) = P_{\rm c}(s)E_{\rm c}(s)\,.
\label{summaryP}
\ee
Here $P_{\rm c}(s)$ is the survival probability for completely correlated
walks, given by \eqn{increasingB}, in which $S_{\rm crit}$ is finite
only when the barrier increases faster than $\sim \sqrt{s}$, and in
particular $S_{\rm crit}\to\infty$ for constant and decreasing
barriers. The correction factor $E_{\rm c}(s)$ is given by
\be
E_{\rm c}(s)=\exp\left(\int_0^s \frac{{\rm d}s^\prime}{\Gamma
   s^\prime}\, \ln c(<B(s^\prime)|s^\prime)\right)\,,
\label{summaryE}
\ee
where $c(<\del|s)$ was defined in \eqn{cdef}.

Although this extension to moving barriers is relatively 
straightforward -- it is far simpler than for walks with uncorrelated 
steps, or for the field theoretic approach -- our comparison with 
Monte-Carlo simulations indicates that there is room for improvement.  
In particular, Figure~\ref{vfvLinear} showed that, for barriers which 
decrease sufficiently rapidly with $s$ (for linear barriers, 
this is true for negative enough values of $\beta$), the solution 
associated with completely correlated walks is an excellent 
approximation.  I.e., the correction factor $E_{\rm c}(s)$ should 
simply be set to unity.  It is easy to see why this happens:  for 
barriers which fall steeply -- where steep means $B(s)$ changes 
height by more than $\sqrt{s}$ over the correlation length scale 
$s\Gamma$ -- the barrier is almost vertical, and so one need not 
worry about walks which cross the barrier more than once.  In this 
sense, the walks behave as though they are completely correlated.
We are currently investigating whether our simple extension of the
Peacock-Heavens ansatz can be improved to make $E_{\rm c}$ depend, 
not just on the correlation length scale $\Gamma$, but on how $\Gamma$ 
relates to the scale barrier shape.  

In Section~\ref{constrained}, we described a simple ansatz which allows 
one to transform the first crossing distribution of unconditional walks 
into a rather good estimate of the first crossing distribution associated 
with walks which are constrained to pass through a certain point in the 
$(\delta,S)$-plane before crossing the barrier.  
When applied to the Peacock-Heavens approximation, our ansatz for 
the conditional distribution (equations~\ref{cCond} and~\ref{v10}) 
works rather well (Figure~\ref{vfvCond}).  As models for the 
unconditional distribution improve, we expect our ansatz to continue 
to provide a useful approximation, because the numerical solution 
does appear to scale as predicted (Figure~\ref{y10fy10}).  

The simplicity of our approach to the problem of conditioned walks 
derives from approximating the problem, which potentially involves 
$n$-point distributions, to bivariate distributions.  To impose more 
than one constraint one must simply modify equations~(\ref{cCond}) 
and~(\ref{v10}); this is a straightforward generalization that we did 
not explore further, but is clearly a straightforward way to 
incorporate Assembly bias effects of the sort identified by 
Sheth \& Tormen (2004), and since studied by a number of authors.  

We were careful to state at the beginning that our primary interest 
is in how well one can describe the first crossing distribution when 
steps are correlated. This is because, although our results provide 
increased understanding of halo abundances and evolution, a number of 
issues must be addressed before they can be used to provide quantitative 
constraints on cosmological parameters.  

First, the correlated walk problem is known to underpredict the 
abundances of clusters (Bond \etal\ 1991).  
There are at least three possible resolutions:  
i) the spherical model for collapse dynamics is wrong -- $\delta_c$ must 
   be smaller, or factors other than the initial overdensity matter 
   (following, e.g., Sheth \etal\ 2001 for clusters);
ii) there is substantial scatter around the actual value of $\delta_c$ -- 
   in which case one must decide whether the appropriate solution is 
   to make the barriers associated with the void-in-void and void-in-cloud 
   problems fuzzy or stochastic (following Sheth \etal\ 2001, or 
   Maggiore \& Riotto 2010b, respectively, and also see discussion 
   in Appendix~B of Sheth \& Tormen 2002), or whether it is better to 
   simply convolve our solution for fixed $\delta_c$ with a 
   distribution $p(\delta_c)$ of values (e.g. Appendix~\ref{stoc}); 
iii) in addition to accounting for correlated steps, we must account for 
   the fact that the appropriate ensemble of walks over which to average 
   also contains correlations -- in effect, our calculation has assumed 
   each walk is a potential halo center, whereas only a few walks really 
   are (compare Figures~2 and~3 in Sheth \etal\ 2001).  
For points ii) and iii) at least, this means that the fundamental 
assumption of the approach, equation~(\ref{ansatz}), is incorrect. 

In addition, the solution to the two barrier problem associated for walks 
with uncorrelated steps is known to give a better description of halo 
merger histories than is the one for correlated steps (Bond \etal\ 1991).  
Thus, even though we have shown how to extend the Peacock-Heavens 
approach so as to provide a reasonable description of the random walk 
problem, we will now have to account for at least one of the three 
issues discussed above, point iii) in particular, before we can claim 
to have a better model for halo formation.  
In the present case, this problem is two-fold, since it is easy to see
that the Peacock-Heavens approximation will break when the barrier to 
be crossed is low, since then the assumption that the walk can travel 
a distance $s\Gamma$ without crossing the barrier is no longer accurate.  
Further study along these lines is in progress.

\section*{Acknowledgments}
TYL was supported by World Premier International Research Center
Initiative (WPI Initiative), MEXT, Japan and JSPS travel grant
(Institutional Program for Young Researcher Overseas Visits), 
and is grateful to the Center for Particle Cosmology at the University
of  Pennsylvania for hospitality. RKS is supported in part by NSF-AST
0908241.

\appendix

\section{Approximation}\label{phMatrix}
\noindent
For a random walk with $n$ discrete time steps $\{s_i\}$, the joint
probability distribution for the sequence of heights $\{ \hat\del_i\}$
is a multivariate Gaussian with covariance matrix
$C_{ij}=\avg{\hat\del_i\hat\del_j}$.  The probability that, after each
of the $n$ steps, the height of the walk lies below $\delta_c$, is
given by:  
\begin{align}
P(s_n) &= \int_{-\infty}^{\delta_c} \frac{d\del_1}{\sqrt{2\pi}} \cdots
 \int_{-\infty}^{\delta_c} \frac{d\del_n}{\sqrt{2\pi}}\,\frac1{|{\rm
     det} C|^{1/2}} \exp\left[-\frac12 \del^T C^{-1} \del\right]
\nonumber\\
&=\int_{-\infty}^{\delta_c}d\del_1 \cdots
\int_{-\infty}^{\delta_c}d\del_n \nonumber\\
&\ph{\int_{-\infty}^{\delta_c}}\times
\lamint{1}\cdots\lamint{n}\exp \left[i\lam^T\del -\frac12
  \lam^TC\lam\right] \,,
\label{app-surv-exact}
\end{align}
where the superscript $T$ denotes a transpose of an $n$-dimensional
vector. Rescaling the variables using
 $\del_i\to x_i=\del_i/\sqrt{s_i}$; $\lam_i\to y_i=\lam_i\sqrt{s_i}$, 
and defining the normalised covariance matrix
 $r_{ij} = C_{ij}/\sqrt{s_is_j}$ 
and scaled barrier heights $\nu_i=\del_c/\sqrt{s_i}$, this survival
probability becomes  
\begin{align}
P(s_n) &= \int_{-\infty}^{\nu_1}dx_1 \cdots
\int_{-\infty}^{\nu_n}dx_n \nonumber\\
&\ph{\int_{-\infty}^{\nu_1}}\times
\int_{-\infty}^{\infty}\frac{dy_1}{2\pi} \cdots
\int_{-\infty}^{\infty}\frac{dy_n}{2\pi} \exp \left[iy^Tx-\frac12 y^Tr y\right] \,.
\label{app-surv-rescale}
\end{align}
In this language, the completely correlated limit is equivalent to
setting $r_{ij}=1$ for all $(i,j)$, since this will result in the joint
probability distribution being a univariate Gaussian in say $x_1$,
multiplying a product of Dirac delta's $\del_D(x_j-x_1)$. This will
lead to precisely the survival probability discussed in Section~\ref{nuwalks}. 

This language also allows us to understand the Peacock-Heavens
ansatz. One can easily check that the discretized survival probability 
in \eqn{PHsurvdiscrete} follows by choosing a sequence of $N<n$ steps
$\{s^{(I)}\}$ and setting $r_{ij}$ to the block-diagonal form 
\be
r = \left( 
\begin{array}{c|cc|ccc|cc}
r^{(1)}&&0&&0&&&0\\ \hline
&&&&&&&\\
0&&r^{(2)}&&0&&&0\\ 
&&&&&&\\ \hline 
&&&&&&&\\ 
0&&0&&r^{(3)}&&&0\\ 
&&&&&&&\\ \hline
&&&&&&\cdot&\\
0&&0&&0&&&\cdot
\end{array}
\right)
\label{app-cc-corrmat}
\ee
with $N$ blocks $\{r^{(I)}\}$ of length $(s_I-s_{I-1})$,  each
corresponding to a set of completely correlated steps :
$r^{(I)}_{ij}=1$ for all $(i,j)$ in the $I^{\rm th}$ block. 
The Peacock-Heavens prescription tells us to choose the sequence
$\{s^{(I)}\}$ to be logarithmically equi-spaced with
 $(\Del s)_I\sim \Gamma s^{(I)}$.  
This clarifies why the Peacock-Heavens ansatz is only an approximation, 
since the true structure of $r_{ij}$ is different.  E.g., for Gaussian 
smoothing with $n=-1$ we have $r_{ij} = 2\sqrt{s_is_j}/(s_i+s_j)$, which 
is of course unity along the diagonal, but falls off gradually rather 
than in sharp jumps. Nevertheless, as the main text shows, the 
subsequent steps in the ansatz (namely, the continuum limit and the 
evaluation of $\Gamma$) lead to a remarkably accurate prescription for 
the first crossing distribution.  

\section{TopHat correlation parameter}\label{Wth}
\noindent
The Peacock-Heavens correlation parameter $\Gamma$ for an arbitrary
filter and power spectrum, follows from equations (3.24) and (3.15b)
of Bond \etal\ (1991). The case of the Gaussian filter and generic
power spectra was discussed in the main text. For a generic filter and
power law power spectrum $P(k)\propto k^n$, $\Gamma$ is given by
\begin{align}
\Gamma&=2\pi\ln(2)\nonumber\\
&\times
\left[\frac{\left(\int_0^\infty dx
    x^{n+2}W(x)^2\right)\left(\int_0^\infty dx
    x^{n+4}W^\prime(x)^2\right)}{\left(\int_0^\infty dx
    x^{n+3}W(x)W^\prime(x)\right)^2} -1\right]^{-1/2}\,,
\label{app-Gamrewrite}
\end{align}
where $W(x)=W(kR)$ is the smoothing filter, and a prime denotes a
derivative with respect to the argument. For the TopHat we have 
\be
W(x) = \frac3x j_1(x) ~~;~~ W^\prime(x) = \frac3x \left(j_0(x) -
\frac3x j_1(x)\right)\,,
\label{app-THfilter}
\ee
where $j_0(x)=\sin(x)/x$ and $j_1(x)=(\sin(x) - x\cos(x))/x^2$ are
spherical Bessel functions of the first kind. It is useful to define
the integrals 
\begin{align}
I_{00} &= \int_0^\infty dx x^{n+2}j_0(x)^2 \,, \nonumber\\ 
I_{01} &= \int_0^\infty dx x^{n+1}j_0(x)j_1(x) \,,\nonumber\\ 
 I_{11} &= \int_0^\infty dx x^{n}j_1(x)^2\,, 
\label{app-besselintegs}
\end{align}
in terms of which we can write $\Gamma$ for the TopHat filter as
\be
\Gamma = 2\pi\ln(2)
\frac{\left|I_{01}-3I_{11}\right|}{\sqrt{I_{00}I_{11} - I_{01}^2}}
\,. 
\label{app-GamTHbesselintegs}
\ee
The integrals in \eqn{app-besselintegs} do not converge for all values
of $n$. In particular, we have
\begin{align}
I_{00} &= 2^{-n}\sin(n\pi/2)\Gamma(n-1)[n(n-1)/4] ~~,~-3<n<-1\,, \nonumber\\
I_{01} &= -2^{-n}\sin(n\pi/2)\Gamma(n-1)[(n+1)/2] ~~,~-3<n<1\,, \nonumber\\
I_{11} &= 2^{-n}\sin(n\pi/2)\Gamma(n-1)[(n+1)/(n-3)] ~~,~-3<n<1\,,
\label{app-besselintegseval}
\end{align}
where $\Gamma(z)$ is the Euler gamma function. 

The Peacock-Heavens correlation parameter is therefore defined for
$-3<n<-1$, and can be simplified to give
\be
\Gamma = 2\pi\ln(2) \sqrt{\frac{(n+1)(n+3)}{(n-3)}} ~~,~ -3<n<-1\,.
\label{app-GamTHfinal}
\ee
The TopHat filter also leads to results slightly different from the
Gaussian, for the conditioned walks discussed in Section 4, in
particular for the form of $r$. For example, if we set $x\equiv
R_s/R_S \le 1$ then, for the TopHat filter, $r = \sqrt{x}(5 - x^2)/4$
if $n=-2$, so  $S_\times/S \to 5/4$ as $S/s\to 0$. If $n=-1$, then 
 $r = \{2x(1 + x^2) + (1-x^2)^2\,\log[(1-x)/(1+x)]\}/(4x^2)$, 
so $S_\times/S \to 4/3$ as $S/s\to 0$. And if $n=0$, then the
real-space TopHat is just like a sharp k-space filter, so
$S_\times/S\to 1$.  Thus, in general, the numerical coefficient is
smaller than when the filter is Gaussian.

\section{Stochastic barriers}\label{stoc}
Treating the problem of halo formation by a single deterministic 
barrier is at best a crude approximation.  In triaxial collapse 
models, themselves crude approximations, the collapse barrier is a 
function of three variables, the joint distribution of which depends 
on scale (Sheth, Mo \& Tormen 2001).  In excursion set language, at 
each step $n$, one asks if $\delta_n > \delta_{\rm ec}(e_n,p_n)$ 
(recall that $n$ is monotonically related to the smoothing scale).  
Since $e_n$ and $p_n$ are random numbers (they are neither Gaussian 
distributed, nor independent of $\delta_n$), one may think of 
$\delta_{\rm ec}$ as being a function of $n$, so the problem is now
that of a random walk ($\delta_n$) crossing a barrier ($\delta_{\rm ec}$) 
whose height is stochastic function of $n$.  If projected onto the 
$(\delta,S)$ plane, this translates to scatter in the value of the 
critical density required for collapse and $S$ (e.g. Figure~B1 in 
Sheth \& Tormen 2002).  

Since the stochasticity of the barrier is related to that of $e_n$ 
and $p_n$, and these have variance proportional to that of $\delta_n$, 
one expects the variance of the barrier height to be proportional to 
$S$.  If one further assumes that changes in the barrier height are 
uncorrelated with changes in $\delta$, and that the steps in the 
barrier height are drawn from a Gaussian distribution (even though 
steps in $e$ and $p$ are not), then, for uncorrelated steps, the net 
effect of the stochastic barrier is simply to rescale all variances:  
$S\to S/a$ for some $a<1$.  This is most easily seen by noting that 
now one asks if $\delta_n > \delta_{\rm c} + b_n$, where $b_n$ is a 
Gaussian number with variance $B_n = DS_n$, for some $D>0$.  
Since $\delta$ and $b$ are both Gaussian distributed, requiring 
$\delta_n-b_n > \delta_{\rm c}$ is the same as requiring $g_n$, a 
Gaussian number with variance $S_n + B_n = S_n(1+D)$, exceed 
$\delta_{\rm c}$.  This rescaling of the variance means that the 
problem is the same as considering when the previous sum of $n$ 
Gaussian numbers first exceeds $\delta_{\rm c}/\sqrt{1+D}$.  
This is attractive because just such a rescaling appears to be 
necessary to reconcile the shape of the first crossing distribution 
with the measured counts of halos in simulations:  
e.g., Sheth \& Tormen (1999) suggest that $D\approx 0.4$.  
However, the rescaling of the variance which is most naturally 
associated with triaxial collapse models, $1+D$ is not as large as 
the factor of $1/0.7$ that is needed (Sheth, Mo \& Tormen 2001), if the 
fundamental ansatz of equation~(\ref{ansatz}) is correct 
(we noted in the final discussion section of the main text that 
Figures 2 and 3 of Sheth et al. 2001 suggest it is not).  

For correlated steps the issue is more complicated, since one must 
now decide if the same smoothing process which correlates the steps 
in $\delta$ also correlates the steps in the barrier height (and also:
if the steps in $\delta$ were previously independent of those in the 
barrier height, does the smoothing now correlate them?).  In triaxial 
collapse models this is indeed the case ($e$ and $p$ are defined in 
the same physical volume as $\delta$).  Hence, the steps in the barrier 
are correlated in the same way as for $\delta$, so the net effect is 
again to simply rescale variances.  In particular, the Peacock-Heavens 
correlation parameter $\Gamma$ is not modified.  

On the other hand, if the stochasticity in the barrier height is due 
to processes on another scale, or to processes which have a different 
correlation structure than $\delta$, then $\Gamma$ will be modified.  
E.g., if the steps in barrier height are uncorrelated even when those 
in $\delta$ are, then $\Gamma \to \Gamma(1+D)$ where $D$ is related 
to the variance in the barrier height (i.e., it also determines the 
rescaling of $S$).  This is exactly analogous to the rescaling 
$\kappa\to \kappa/(1+D)$ which Maggiore \& Riotto assume in their 
treatment of stochasticity. In a paper which appeared after an earlier
version of the present work, Corasaniti \& Achitouv (2011b) use this
argument in the treatment of random walks and proceed to fit the
resulting first crossing distribution to a mass function
from $N$-body simulations using \eqn{ansatz}, leaving $D$ (and 
the barrier slope) as free parameters and finding agreement at the 5\%
level.  
We emphasize, however, that this rescaling is effectively invoking
some unspecified process which operates on a scale that is not the
same as that on which $\delta$ was defined. 
More importantly, since the ansatz in \eqn{ansatz} is most likely
incorrect (see above), conclusions about how reasonable the fitted
value of $D$ is are suspect, given the obvious differences between
Figures 2 and 3 in Sheth, Mo \& Tormen (2001).
And finally, this treatment is motivated by the
observed scatter in the value of \delc\ -- rather than on
theoretical considerations of triaxial collapse -- which introduces a
new ambiguity as we discuss next.  

\subsection{Distribution of deterministic barriers}
The discussion above noted that if the collapse barrier is a 
stochastic function of scale, then this will appear as scatter in 
the critical density at fixed $S$.  But it is important to note that 
the converse is not true:  the observation of scatter in the critical 
density at fixed $S$ (e.g. Figures~2 and~3 of Sheth et al. 2001) does 
{\em not} imply that the barrier itself is stochastic.  For example, 
one might find scatter in $\delta_{\rm c}$ at fixed $S$ if there were 
simply a distribution of barrier shapes, each one of which was 
deterministic.  To illustrate this (trivial) point, we consider three 
simple examples below.  In all cases we illustrate our arguments 
using walks with uncorrelated steps, but note that the 
extension to correlated steps presents no conceptual difference.  

\subsubsection{Constant barriers}
The first assumes that the first crossing problem of interest is always 
that of a constant barrier; however, the barrier height is different 
for different walks.  In this case, the distribution of interest is 
simply 
\begin{equation}
 Sf(S) = \int {\rm d}\delta_{\rm c}\,p(\delta_{\rm c})\,Sf(S|\delta_{\rm c}).
\end{equation}
If we use 
\begin{equation}
 p(\delta_{\rm c})\,{\rm d}\delta_{\rm c} = 
 \frac{{\rm d}\delta_{\rm c}}{\Delta_{\rm c}}\,
 \frac{4\pi \delta_{\rm c}^2}{\Delta_{\rm c}^2}
 \frac{\exp(-\delta_{\rm c}^2/2\Delta_{\rm c}^2)}{(2\pi)^{3/2}}
\end{equation} 
to model the distribution of $\delta_{\rm c}$ around a typical value 
$\Delta_{\rm c}$, then the integral above can be done analytically:
\begin{equation}
 Sf(S)
  = \frac{2}{\pi} \frac{\Delta_{\rm c}/\sqrt{S}}{(1 + \Delta_{\rm c}^2/S)^2}.
\end{equation}
This shows that the small $S\ll\Delta_{\rm c}$ tail is modified 
dramatically, from an exponential to a power-law.  
If $f(s)$ is indeed related to halo counts via equation~(\ref{ansatz}), 
then the exponentially falling tail of halo counts argues against 
this as being the relevant model of stochasticity.  

\subsubsection{Linear barriers}
The model above has the same distribution of critical barrier heights 
at all $S$.  A simple model which allows $S$-dependent scatter is to 
assume that the deterministic barrier is linear, 
 $\delta_{\rm c} + \beta S$, but that different walks have different 
values of $\beta$.  (This has some merit, since $S$-dependent scatter 
is suggested by Figures~2 and~3 of Sheth et al. 2001.)  
If we use a Gaussian distribution for $\beta$, then 
\begin{eqnarray}
 Sf(S) &=& \int {\rm d}\beta\,p(\beta)\,Sf(S|\beta) \nonumber\\
   &=& \int {\rm d}\beta\,
       \frac{{\rm e}^{-\beta^2/2\Sigma_\beta^2}}{\sqrt{2\pi\Sigma_\beta^2}}\,
       \frac{\delta_{\rm c}}{\sqrt{S}}\,\frac{{\rm e}^{-(\delta_{\rm c} + \beta S)^2/2S}}
                                            {\sqrt{2\pi}}\nonumber\\
   &=& \frac{{\rm e}^{\delta_{\rm c}^2\Sigma_\beta^2/2(S\Sigma_\beta^2 + 1)}}
        {\sqrt{1 + \Sigma_\beta^2 S)}}\,
       \frac{\delta_{\rm c}\,{\rm e}^{-\delta_{\rm c}^2/2S}}{\sqrt{2\pi S}}\nonumber\\
   &=& \frac{\delta_{\rm c}}{\sqrt{S(1 + \Sigma_\beta^2 S)}}\,
       \frac{{\rm e}^{-\delta_{\rm c}^2/2S(1 + S\Sigma_\beta^2)}}{\sqrt{2\pi}}.
\end{eqnarray}
When $\Sigma_\beta=0$ then this reduces to the solution for a constant 
barrier.  So the effect of stochasticity is almost like rescaling 
$S\to S(1 + \Sigma_\beta^2 S)$.  In this case, the exponential 
cut-off at small $S$ is not modified, but the distribution at 
large $S$ is changed dramatically.  The first term in the penultimate 
expression may be thought of as a correction to the deterministic 
$\beta=0$ case.  This factor is $\exp(\delta_{\rm c}^2\Sigma_\beta^2/2)$ 
when $S\ll \Sigma_\beta^2$, but it decreases rapidly at large $S$.  
Again, if equation~(\ref{ansatz}) is correct, then the enhanced 
counts at small $S$ is encouraging, but $\Sigma_\beta$ may have to 
be rather large for this to be a competitive model.  Of course, one 
is free to explore other models for $p(\beta)$.  We will not do 
so here, because we feel we have made our main point -- a distribution 
of deterministic barriers produces a distribution of critical densities 
at fixed $S$, and this distribution affects the quantity which enters 
equation~(\ref{ansatz}).  

\subsubsection{Square-root barriers}
Before closing our discussion, we think it is worth showing explicitly 
that, by choosing the barrier shapes and their distribution 
appropriately, one can obtain effective first crossing distributions 
which closely approximate that for a single deterministic barrier with 
rescaled height (or, alternatively, with rescaled variance).  
This happens to be true if the deterministic barriers have height
 $\delta_{\rm c} + \beta \sqrt{S}$, 
with a Gaussian distribution of $\beta$.  This case is tractable, 
because the first crossing distribution of a square-root barrier 
(of specified $\beta$) is known.  Although the exact solution is 
complicated (Breiman 1966), it is quite well approximated by 
\begin{equation}
 Sf(S|\beta) \approx 
        \left(1 + \frac{\beta}{4}\frac{\sqrt{S}}{\delta_{\rm c}}\right) 
               \frac{\delta_{\rm c}}{\sqrt{S}}\,
               \frac{{\rm e}^{-(\delta_{\rm c} + \beta \sqrt{S})^2/2S}}{\sqrt{2\pi}}
\end{equation}
(Sheth \& Tormen 2002; Moreno et al. 2008), so that 
\begin{eqnarray}
 Sf(S) &=& \int {\rm d}\beta\,
           \frac{{\rm e}^{-\beta^2/2\Sigma_\beta^2}}{\sqrt{2\pi\Sigma_\beta^2}}\,
            Sf(S|\beta) \\
   &=& \left(1 - \frac{\Sigma_\beta^2}{4(1 + \Sigma_\beta^2)}\right)
       \frac{\delta_{\rm c}}{\sqrt{S(1 + \Sigma_\beta^2)}}\,
       \frac{{\rm e}^{-\delta_{\rm c}^2/2S(1 + \Sigma_\beta^2)}}{\sqrt{2\pi}}. \nonumber
\end{eqnarray}
This final expression is remarkable, because it shows that $Sf(S)$ has 
the same shape as the first crossing distribution of a barrier of 
constant height: 
i.e., $\beta=0$ and $\delta_{\rm c}/\sqrt{1 + \Sigma_\beta^2}$.  
The constant of proportionality differs from unity, presumably 
because the approximation we used for the crossing of a square-root 
barrier is not quite normalized.  
But even for $\Sigma_\beta = 0.65$, for which the effective height is 
the desired $\sqrt{0.7}\delta_{\rm c}$, this normalization factor is 
only $0.925$.  Therefore, we have not bothered to repeat this exercise 
using Breiman's exact expression for the first crossing distribution.  

\subsubsection{The potential to test} 
We conclude that such models for the stochasticity are potentially 
rather interesting.  Perhaps more importantly, they are testable.  
By following essentially the same steps taken by Sheth et al. (2001) 
to produce their Figures~2 and~3, it is rather straightforward to test 
if the stochasticity in $\delta_{\rm c}$ at fixed mass is indeed due to 
a distribution of deterministic barriers.  The same is not true if the 
barrier height were a stochastic function of scale. Until such tests
are performed, we argue that the question of how to correctly describe
the observed scatter in barrier heights at fixed mass remains an open
one. We leave this to future work.  

\label{lastpage}

\end{document}